\documentclass[12pt]{article}
\usepackage{graphicx}
\newcommand{\be}{\begin{equation}}
\newcommand{\ee}{\end{equation}}
\begin{document}

{} \hfill {\bf \large IFT/07/02}

\vskip1in
\centerline{\Large {\bf  Limit cycles in quantum theories}}
\vskip .3in
\centerline{\bf  Stanis{\l}aw D. G{\l}azek 
\footnote[1]{Work supported in part by KBN Grant No. 2 P03B 016 18.}}
\vskip .05in
\centerline{\small Institute of Theoretical Physics, Warsaw University}
\centerline{\small ul.  Ho{\.z}a 69, 00-681 Warsaw}
\vskip.05in
\centerline{and}
\vskip.05in
\centerline{\bf  Kenneth G. Wilson}
\vskip .1in
\centerline{\small Department of Physics, The Ohio State University}
\centerline{\small 174 West 18th Ave., Columbus, Ohio 43210-1106}
\vskip.5in
\begin{center}
{\bf Abstract}
\vskip.1in
\parbox{13cm}{\small
Renormalization group limit cycles may be a commonplace for 
quantum Hamiltonians requiring renormalization, in contrast 
to experience to date with classical models of critical points, 
where fixed points are far more common. We discuss the simplest 
model Hamiltonian identified to date that exhibits a renormalization 
group limit cycle. The model is a discrete Hamiltonian with two 
coupling constants and a non-perturbative renormalization group 
that involves changes in only one of these couplings and is soluble 
analytically. The Hamiltonian is the discrete analog to a continuum 
Hamiltonian previously proposed by us.}  
\end{center}
\vskip1in

\section{Introduction}
\label{sec:intro}

In 1971, one of us suggested that renormalization group equations 
could have limit cycle solutions as well as fixed points when the 
renormalization group equations involve two or more coupling constants 
\cite{Wlc}. The 1971 paper did not mention the possibility that 
renormalization group equations might have limit cycle solutions 
even for differential equations for only one coupling constant. 
But in 1993, the two of us defined a simple Hamiltonian that requires 
the renormalization of just a single bare coupling constant $g_\Lambda$, 
where $\Lambda$ is the cutoff \cite{Glazek:1993qs}. The bare coupling 
constant, we demonstrated, exhibits limit cycle behavior in 
the limit $\Lambda \rightarrow \infty$. To be precise the coupling 
constant $g_\Lambda$ was found to decrease steadily as $\Lambda$ increased, 
until $g_\Lambda$ reached $-\infty$, after which it jumped to $+\infty$ 
and started its next cycle of steady decrease. However, we did not 
recognize or comment that this behavior constituted a limit cycle. 
Moreover, the Hamiltonian we defined is singular, in such a way that 
it presumably does not have well-defined s-wave scattering phase 
shifts. The model has well-defined bound states, and our renormalization 
group analysis was based on keeping the finite bound state energies 
fixed as $\Lambda \rightarrow \infty$.

The model Hamiltonian we published in 1993 has no known physical
application. But in 1999, Bedaque, Hammer, and Van Kolck showed 
that a three-body Hamiltonian with two- and three-body delta function 
potentials and a cutoff in momentum space is renormalizable and that 
the three-body coupling approaches a limit cycle as the cutoff lambda  
approaches $\infty$ \cite{BHK}, much as it does in our 1993 model. 
Bedaque et al built on earlier work of Thomas \cite{Thomas} and 
Efimov \cite{Efimov}, that was recently reviewed by Nielsen, Fedorov, 
Jensen, and Garrido \cite{NFJG}. However, the mathematics of the three 
body system is complex, and it is surely useful to have simpler models 
to study that exhibit limit cycles too.

In this paper, we discuss a discretized version of the Hamiltonian 
we introduced in 1993. With a cutoff, the discretized model takes 
the form of a finite size matrix with discrete eigenvalues. The 
discrete matrix has a diagonal sub-matrix, plus two off-diagonal 
pieces with two coupling constants. The Hamiltonian requires 
renormalization in the limit of infinite cutoff. Just as in the 
continuum case, the renormalization can be constructed analytically 
and leads to a limit cycle in one of the two couplings, while the 
second coupling constant stays fixed. But the analytically obtained 
limit cycle is defined only for a discrete sequence of cutoffs, 
rather than for a continuously varying cutoff. Using numerical
procedures, it is possible to define and compute a renormalization 
process with a continuously varying cutoff, but our studies with 
the continuous cutoff variation will not be discussed here.

The finite matrix Hamiltonians can be diagonalized numerically, when 
the cutoff is small enough. The renormalizability of the Hamiltonians 
we will discuss here can be demonstrated to high numerical accuracy 
with cutoffs small enough to allow numerical diagonalization. We will 
provide the demonstration with a comparison of the eigenvalues of two 
matrices with two different cutoffs, one of size 37x37, the other of 
size 42x42.

The Hamiltonian to be used in this paper has the form
\be
H_{mn}(g_N,h_N) = (E_m E_n)^{1/2}\left[\delta_{mn} - g_N 
- i h_N s_{mn}\right] \; ,
\label{hN}
\ee
where $m$ and $n$ are integers. For $m=n$, $\delta_{mn}= 1$ and 
$s_{mn} = 0$. For $m \neq n$, $\delta_{mn}=0$ and $s_{mn} = (m-n)/|m-n|$.
The numbers $E_n = b^n$ with $b>1$, are eigenvalues of the operator 
$H_0$ that has matrix elements $\langle m|H_0|n\rangle = H_{mn}(0,0)$.
The eigenvalues are called kinetic energies of the corresponding 
eigenstates, $|n\rangle$. The remaining part of the Hamiltonian, $H_I = 
H - H_0$, is called an interaction, and $H_I(0,0)=0$. The largest energy 
allowed in the dynamics is $\Lambda_N = b^N$, which defines the 
ultraviolet cutoff so that the subscripts $m, n \le N$. 

The continuous version of this model \cite{Glazek:1993qs} is recovered
in the limit $b\rightarrow 1$. The discrete model itself has been 
discussed in the case $h_N=0$ using similarity renormalization 
group idea \cite{GlazekWilson12} and Wegner's equation \cite{GWW}. 
It was shown that the model exhibits asymptotic freedom for $h_N=0$. 
Hamiltonians of Eq. (\ref{hN}) with $h_N=0$ can be derived in a number 
of ways, ranging from a discretization of a nonrelativistic Schr\"odinger 
equation for a particle on a plane with a two-dimensional $\delta$-potential, 
to a discrete version of the transverse dynamics of partons in quantum 
field theory. 

At first, the model with $h_N \neq 0$ does not appear much different 
from the one with $h_N=0$. All Hamiltonians defined by Eq. (\ref{hN}) 
are hermitian and have a general ultraviolet logarithmically 
divergent structure. The divergence originates in the fact that the 
far off-energy-shell matrix elements of $H_{mn}$, i.e. those with 
$|m-n| \gg 1$, behave like $(E_m E_n)^{1/2}$. Therefore, perturbation 
theory produces large-energy contributions of the form $\sum_i 
(E_m E_i)^{1/2}\,(1/E_i)\,(E_i E_n)^{1/2}$, and every energy scale 
contributes a significant amount to the sum, which grows linearly with 
the number of initially included scales, $N = \ln \Lambda_N / \ln b$. 
The dependence of the unrenormalized coupling constants $g_N$ and $h_N$ 
on $N$, is expected to compensate this effect. As we will demonstrate 
below, when $h_N$  is not zero, the model exhibits limit cycle behavior 
as N goes to infinity. We will prove this by deriving a renormalization 
group equation that determines $g_{N-1}$ (used with a cutoff 
$\Lambda_{N-1}$ ), given $g_N$  and $h_N$  used with cutoff $\Lambda_N$, 
such that the low energy eigenvalues stay fixed. 

This paper is organized as follows. Section \ref{sec:rgt} describes 
the discrete RG flow of the model Hamiltonians $H(g_N,h_N)$ with $N$. 
The structure of eigenstates of these Hamiltonians is also briefly 
described. Section \ref{sec:ne} shows a numerical example of the 
spectrum of renormalized Hamiltonians and their discrete rescaling 
properties. The spectrum is found to contain a sequence of negative 
(bound state) eigenvalues, one per cycle. Section \ref{sec:c} 
concludes the paper.

\section{Renormalization group trajectories}
\label{sec:rgt}

The eigenvalue problem
\be
\sum_{n= -\infty}^N H_{mn}\psi_n = E \psi_m \; ,
\label{eve}
\ee
can be solved for $\psi_m$, $m \le N$, assuming that 
one knows $E$, by using the Gaussian elimination procedure. 
In the first step one can express $\psi_N$ in terms
of all other components, $\psi_n$ with $n < N$.
In the next step, one expresses $\psi_{N-1}$ in terms 
of components $\psi_n$ with $n < N-1$, and so on. 
In fact, this procedure can be carried out without 
knowing $E$ if the eliminated energy scales are much 
larger than $E$. Thus, when one is interested in the 
eigenvalues of order $b^M$ with $M \rightarrow -\infty$, 
one can eliminate states from some huge energy range 
between $\Lambda_N = b^N$ and some fixed $\Lambda_0 = 
b^{N_0}$, where $N_0$ is much smaller than $N$ and 
still $\Lambda_0 \gg E$. The new, much smaller eigenvalue 
problem will have the cutoff $\Lambda_0$ and the 
corresponding coupling constants. When the recursion for 
the whole sequence of the coupling constants $(g_N,h_N)$, 
$(g_{N-1},h_{N-1})$,..., $(g_{N_0},h_{N_0})$, is found, 
one can trace it back starting from some assumed finite 
values $(g_0,h_0)$ at a finite $N_0$ and see what happens 
with $(g_N,h_N)$ when the cutoff number $N \rightarrow \infty$. 

To carry out the above procedure, it is convenient to 
introduce a small number $\tilde E = E/b^N$ and write 
the eigenstate components in the form $\psi_n = b^{n/2} 
\phi_n$ for all $n \le N$. The eigenvalue Eq. (\ref{eve}) 
becomes:
\newpage
\begin{eqnarray*}
(1-\tilde E)     \phi_N     & = & g_N\sigma_N
+ih_N\sigma_{N-1}\; , \\
(1-b \tilde E)   \phi_{N-1} & = & g_N\sigma_N
+ih_N\left(-\phi_N+\sigma_{N-2}\right) \; , \\
(1-b^2 \tilde E) \phi_{N-2} & = & g_N\sigma_N
+ih_N\left(-\phi_N-\phi_{N-1}+\sigma_{N-3}\right)\; , \\ 
                            & : &  \\
(1-b^n \tilde E) \phi_{N-n} & = & g_N\sigma_N
+ih_N\left(-\phi_N+...-\phi_{N-n+1}+\sigma_{N-n-1}\right)\; , \\
                     & : &  \quad \quad ,
\label{eN}
\end{eqnarray*}
where $\sigma_i = \sum_{j = -\infty}^i \phi_j$. If $E$  
and $n$ are finite and $N\rightarrow \infty$, $b^i \tilde E$
for $i = 0,1,...,n$ on the left hand sides of the above 
equations do not count. The first step of the Gaussian 
procedure gives 
\be 
\phi_N = (g_N+ih_N)\sigma_{N-1}/(1-g_N) \; .
\label{phiN}
\ee
Substituting this result into the remaining equations,
one obtains a new set that does not explicitly involve 
the component with largest kinetic energy but has a 
different coupling constant instead. Namely, 
\begin{eqnarray*}
(1-b \tilde E)   \phi_{N-1} & = & g_{N-1}\sigma_{N-1}
+ih_{N-1}\sigma_{N-2} \; , \\ 
(1-b^2 \tilde E) \phi_{N-2} & = & g_{N-1}\sigma_{N-1}
+ih_{N-1}\left(-\phi_{N-1}+\sigma_{N-3}\right)\; , \\ 
                            & : &  \\
(1-b^n \tilde E) \phi_{N-n} & = & g_{N-1}\sigma_{N-1}
+ih_{N-1}\left(-\phi_{N-1}+...-\phi_{N-n+1}+\sigma_{N-n-1}\right)\; , \\
                            & : & \quad \quad . 
\label{eN-1}
\end{eqnarray*}
The new coupling constant $g_{N-1} = g_N + (g_N^2 + h_N^2)/(1-g_N)$, 
while $h_{N-1} = h_N$ stays unchanged. Making $p$ such steps, 
one transforms $(g_N,h_N)$ into $(g_{N-p},h_{N-p})$, where $h_{N-p} 
= h_N$ and
\be
g_{N-p} = h_N \, \tan\left[\arctan\left({g_N\over h_N}\right) 
+ p \,\arctan(h_N)\right] \; .
\label{sxp}
\ee   
It is visible that $g_{N-p} = g_N$ when $h_N = \tan(\pi/p)$. Equation 
(\ref{sxp}) demonstrates that the coupling constant $g_N$ completes a 
cycle when $\Lambda_N$ is reduced to $\Lambda_N/b^p$, and the cycle 
repeats itself in the further reductions. Fig. 1 shows these cycles 
for $\Lambda_{16} = b^{16}$, $g=0.060606$ \cite{GWW}, and $h_{16} = 
\tan(\pi/5)$. If $g_{16}$ were zero, the cycle would be symmetric with 
respect to the axis $g_N=0$. On the other hand, components of the 
eigenvectors of $H(g_N,h_N)$ satisfy the recursion
\be
\psi_n = b^{1/2}\, {1-ih_N-E/b^{n+1} \over 1+ih_N-E/b^n} \, \psi_{n+1},
\label{psi}
\ee
for all $N$ and $n \le N$ independently of the value of $g_N$. 
\begin{figure}[htb]
\begin{center}
\includegraphics[scale=1.2]{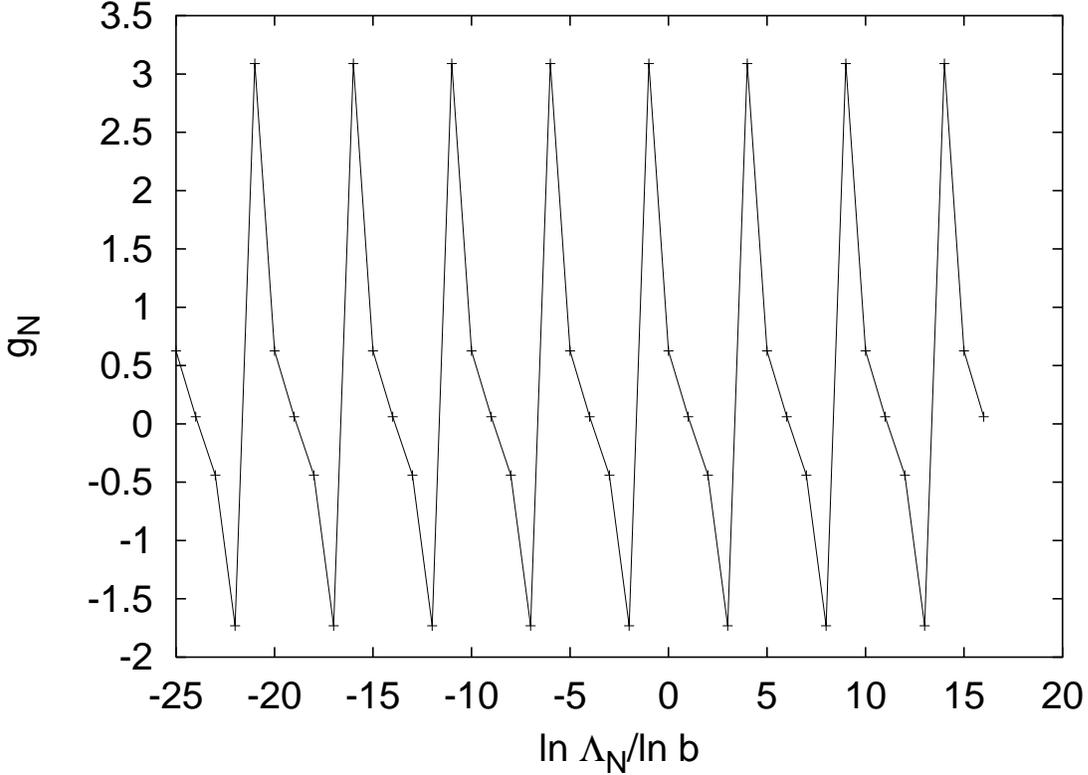}
\end{center}
\caption{The discrete limit cycle dependence of the
coupling constant $g_N$ on the cutoff number $N = \ln \Lambda_N/
\ln b$, $g_{16} = 0.060606$, $h_{16} = \tan(\pi/5)$. The lines are
added between the discrete points only to guide the eye.}
\label{fig1}
\end{figure}

Note that for real Hamiltonian matrices, i.e. 
for $h_N \rightarrow 0$, Eq. (\ref{sxp})implies
\be 
g_N = {g_0 \over 1 + g_0 (N-N_0)} \; ,
\ee
which means asymptotic freedom, or $g_\Lambda = g_0/\left[1+g_0 
\ln (\Lambda/\Lambda_0)/\ln b\right] \rightarrow 0$ when $\Lambda
\rightarrow \infty$. This is a fixed point behavior. In 
contrast, for complex Hamiltonians, Eq. (\ref{sxp}) is equivalent 
to the following renormalization group transformation of $g_\Lambda$ 
from $\Lambda_0$ to $\Lambda$, ($h \equiv h_N = const.$),
\be 
g_\Lambda = h \, { g_0 - h \, \tan \left[c \ln(\Lambda/\Lambda_0)\right]
\over h + g_0  \, \tan \left[c \ln(\Lambda/\Lambda_0)\right]} \; ,
\label{gl}
\ee
where $c = \arctan(h)/\ln (b)$, which exhibits the limit cycle. 

In the continuum limit of $b \rightarrow 1$, the sum over energy 
scales, $\sum_i \rightarrow \int d\Lambda /(\Lambda \ln b)$. 
When the ratio $r_c = \Lambda/\Lambda_0$ of two cutoffs separated
by one complete cycle is kept constant and $b$ tends to 1, the 
difference $p$ between powers $N$ and $N_0$ must grow to infinity, 
as $p =\ln r_c / \ln b$. The corresponding coupling constant 
$h = \tan (\pi/p)$ becomes then equal to $\pi/p = (\pi \ln b)/\ln r_c$ 
and the constant $c$ in Eq. (\ref{gl}) becomes equal to $\pi/\ln r_c$. 
This result shows that the limit $b \rightarrow 1$ in the discrete model 
reproduces the continuum case from Ref. \cite{Glazek:1993qs}, 
and the period ratio is $r_c = \exp (\pi/c)$ for all values
of the imaginary part $c$ of the interaction term. The fixed-point 
behavior with asymptotic freedom is obtained only in the limit
$c \rightarrow 0$, as the RG behavior in a one long cycle. 

\section{Numerical example}
\label{sec:ne}

For illustration of the limit cycle from Fig. 1 and to show the
existence of bound state solutions associated with it, Table 1 
provides results of diagonalization of two Hamiltonian matrices: 
$H_1$ with $m$ and $n$ ranging from $M=-25$ to $N=16$, and $H_2$ 
with $N=11$ and the same $M$. Both Hamiltonians have the same 
values of $b=2$, $h_{16} = h_{11} = \tan(\pi/5) \sim 0.726543$, 
and $g_{16} = g_{11} = 0.060606$ \cite{GWW}. The Hamiltonian $H_2$ 
is considered a result of five discrete RG steps applied to the 
Hamiltonian $H_1$. The infrared cutoff $M$ was introduced in 
order to use a computer for calculations. The lower is $M$ the 
better is the agreement between the example and the cycle theory 
applicable for $M \rightarrow -\infty$. The eigenvalues of small 
modulus (not too small, however, to stay away from the lower
bound $b^M$) are identical in both cases within the displayed
numerical accuracy. For these eigenvalues, the RG analysis 
of the previous section is fully confirmed. For eigenvalues
close to the artificial lower bound $b^M$, the finite size of 
the matrix changes the pattern in a certain way, which is not 
relevant here. 

The scaling phenomenon of interest is described by showing 
the ratios of ratios of successive eigenvalues for one matrix 
to two characteristic numbers. The negative eigenvalues, that 
correspond to bound states ($E_n = b^n$ are positive for all 
values of $n$), appear in a geometric sequence with quotient $32$, 
with accuracy approaching several ppm in the best cases. Thus, 
there is one bound state per cycle among the solutions. The 
positive eigenvalues also appear in a geometric sequence, but 
with a quotient almost equal $2^{5/4}$, which can be understood 
on the basis of dimensional analysis without inspecting details 
of the solution. Namely, after $p$ reduction steps one eliminates 
$p$ states from the dynamics and obtains the Hamiltonian $H_2$ that 
is equivalent to $b^{-p} H_1 = H_1/32$. Since in every cycle among 
the $p$ states only $p-1$ have positive eigenvalues and the
cycle repeats itself indefinitely, the ratio of the successive 
$p-1$ eigenvalues (if such a common quotient exists) must satisfy 
the condition $r^{p-1} = b^p$ in order to scale properly from one 
cycle to another. Therefore, $r= b^{p/(p-1)}$, and in the 
example one obtains the factor $32^{1/4}$. Deviations 
from that rule within a single cycle cannot be understood 
so simply but the calculation shows a repeating sequence 
within a cycle with all elements very close to $r$.  

However, the most prominent feature of the example is the 
appearance of one negative eigenvalue in every cycle. It 
requires further studies to be fully understood. Here it is
noted that when the period $p$ tends to infinity with 
$h_N \rightarrow 0$, one recovers the case from Ref. \cite{GWW}, 
with asymptotic freedom and a bound state, but in infinitely 
many copies connected together in a sequence. In every single 
long-cycle link in that chain, looked at in the direction 
of growing $\Lambda$, the asymptotically free coupling 
decreases logarithmically towards large energies until it 
eventually crosses zero, grows in size and turns up rapidly. 
A new generation bound state is uncovered at that scale, and 
a new cycle starts from a large coupling constant that decreases 
slowly again. 
\begin{center}
{\small
\begin{tabular}{|r|c|c|c|c|}
\hline  
     &                 &           &                 &          \\
 $n$ &$\Lambda= 2^{16}$& $r$       &$\Lambda= 2^{11}$& $r$      \\
\hline
 13  & 0.954734+05     & -         & -               & -        \\
 12  & 0.328953+05     & 0.819481  & -               & -        \\
 11  & 0.131845+05     & 0.953270  & -               & -        \\
 10  & 0.545303+04     & 0.983701  & -               & -        \\
  9  & 0.228087+04     & 0.994830  & 0.298354+04     & -        \\
  8  & 0.956198+03     & 0.997093  & 0.102798+04     & 0.819481 \\
  7  & 0.401063+03     & 0.997590  & 0.412015+03     & 0.953270 \\
  6  & 0.168593+03     & 0.999803  & 0.170407+03     & 0.983701 \\
  5  & 0.709615+02     & 1.001090  & 0.712770+02     & 0.994830 \\
  4  & 0.298253+02     & 0.999652  & 0.298812+02     & 0.997093 \\
  3  & 0.125234+02     & 0.998679  & 0.125332+02     & 0.997590 \\
  2  & 0.526682+01     & 1.000260  & 0.526853+01     & 0.999803 \\
  1  & 0.221724+01     & 1.001270  & 0.221755+01     & 1.001090 \\
  0  & 0.931986+00     & 0.999731  & 0.932040+00     & 0.999652 \\
 -1  & 0.391347+00     & 0.998713  & 0.391357+00     & 0.998679 \\
 -2  & 0.164586+00     & 1.000270  & 0.164588+00     & 1.000260 \\
 -3  & 0.692886-01     & 1.001280  & 0.692889-01     & 1.001270 \\
 -4  & 0.291245-01     & 0.999733  & 0.291245-01     & 0.999731 \\
 -5  & 0.122296-01     & 0.998714  & 0.122296-01     & 0.998713 \\
 -6  & 0.514332-02     & 1.000270  & 0.514332-02     & 1.000270 \\
 -7  & 0.216526-02     & 1.001280  & 0.216526-02     & 1.001280 \\
 -8  & 0.910135-03     & 0.999730  & 0.910135-03     & 0.999730 \\
 -9  & 0.382169-03     & 0.998705  & 0.382169-03     & 0.998705 \\
-10  & 0.160723-03     & 1.000250  & 0.160723-03     & 1.000250 \\
-11  & 0.676587-04     & 1.001230  & 0.676587-04     & 1.001230 \\
-12  & 0.284359-04     & 0.999610  & 0.284359-04     & 0.999610 \\
-13  & 0.119370-04     & 0.998423  & 0.119370-04     & 0.998423 \\
-14  & 0.501683-05     & 0.999594  & 0.501683-05     & 0.999594 \\
-15  & 0.210854-05     & 0.999630  & 0.210854-05     & 0.999630 \\
-16  & 0.882705-06     & 0.995684  & 0.882705-06     & 0.995684 \\
-17  & 0.367003-06     & 0.988875  & 0.367003-06     & 0.988875 \\
-18  & 0.150481-06     & 0.975210  & 0.150481-06     & 0.975210 \\
-19  & 0.584264-07     & 0.923457  & 0.584264-07     & 0.923457 \\
-22  & 0.654169-08     & 0.266298  & 0.654169-08     & 0.266298 \\
 -4  & -.622117-06     & 0.964026  & -.622117-06     & 0.964026 \\
 -3  & -.206506-04     & 0.998881  & -.206506-04     & 0.998880 \\
 -2  & -.661561-03     & 0.999965  & -.661561-03     & 0.999960 \\
 -1  & -.211707-01     & 0.999994  & -.211708-01     & 0.999831 \\
  0  & -.677466-00     & 0.999832  & -.677580+00     & 0.994617 \\
  1  & -.216826+02     & 0.994617  & -.217999+02     & 0.824124 \\
  2  & -.697598+03     & 0.824124  & -.846472+03     & -        \\
  3  & -.270871+05     & -         & -               & -        \\ 
\hline
\end{tabular}}
\vskip.05in
\parbox{16cm}
{\footnotesize{{\bf Table 1.} The columns contain:
$n$ - the approximate integer powers of $2^{5/4}$ ($2^5$) that 
give positive (negative) eigenvalues; $\Lambda = 2^{16}$ - the 
eigenvalues of the Hamiltonian with $N=16$ and $M=-25$ (see the 
text for details); $r$ - the ratio of the ratio of two successive 
eigenvalues to $2^{5/4}$ ($2^5$) for the positive (negative)
eigenvalues; $\Lambda = 2^{11}$ - the eigenvalues of $H$ after
5 discrete RG steps (the same $M$ is used).}}
\end{center}
\section{Conclusion}
\label{sec:c}

The possibility of a long asymptotically free cycle due to
small imaginary couplings, with an abrupt end and a bound
state formation where a new cycle starts, is an interesting 
feature of our specific model. But the central result of our 
paper is that such a simple Hamiltonian exhibits limit cycle 
behavior for non-zero $h$ instead of a fixed point. Together 
with the limit cycle found by Bedaque et al, and implicit in 
earlier work of Efimov and Thomas, this raises the question 
of whether limit cycle behavior is a common feature of 
renormalized quantum theories, outside of those that have 
already been solved and are known to exhibit fixed points.
The model analysis carried out here certainly needs improvements, 
especially concerning details of the bound state formation and 
the sharp turn of the coupling where it approaches 1 and Eq. 
(\ref{sxp}) produces a jump. 

Separately from the existence of the limit cycle, the model is 
also valuable because it can be studied using the similarity 
RG technique (see \cite{GWW}), and the latter can be applied to 
Hamiltonians in quantum field theories of major interest, such 
as QCD (see e.g. \cite{Wilson6,G}). (We do not expect QCD in
particular to show limit cycle behavior: our comments about 
limit cycles apply more to broad classes of Hamiltonians, most 
not yet examined, rather than specific ones already examined as 
extensively as QCD has been.) The authors have generated 
numerical solutions to Wegner's differential equation in the 
model and found continuous similarity RG evolution of effective 
couplings with all key features described here. Thus, when one 
uses the similarity scheme in studies of effective Hamiltonians 
in more advanced models, one may also encounter complex RG 
scenarios. It remains to be seen if the model structure reappears 
in realistic quantum theories.

\end{document}